\documentclass{aa}
\usepackage{graphicx}

\begin{document}

\title{Asymptotic behavior of a stratified perturbation in a three 
dimensional expanding Universe}

\author{D.~Fanelli 
\inst{1} 
\and
E.~Aurell \inst{1,2,3} }

   \offprints{D. Fanelli}

   \institute{Department of Numerical Analysis and Computer Science, KTH, S-100 44 Stockholm, Sweden\\
              \email{fanelli@nada.kth.se}
         \and
            SICS, Box 1263, SE-164 29 Kista, Sweden\\
             \email{eaurell@sics.se}
	 \and
            NORDITA, Blegdamsvej 17, DK-2100 Copenhagen, Denmark
              }

  \date{}

\abstract{
The non-linear evolution of a stratified perturbation 
in a three dimensional expanding Universe is considered. 
A general Lagrangian scheme (Q model) is introduced and
numerical investigations are performed. The asymptotic contraction 
of the core of the agglomeration is studied. A power-law scaling is 
detected and an heuristic interpretation of the numerical findings is 
provided. An asymptotic equation for the multi-stream velocity flow 
is derived and it is shown to agree quantitatively with the dynamics of the 
Q model. The relation to the adhesion model is discussed.
}

\titlerunning{Asymptotic behavior of a stratified perturbation ...}

\maketitle

\section{Introduction} 
\label{s:int}

The present Universe is inhomogeneous with structures of many scales, from
galaxies to galaxy clusters and superclusters. Around
$100000$ years after Big Bang, the Universe was however very nearly 
homogeneous, with density fluctuations of relative magnitude of about 
$1:10^5$.
The presently observed large scale structure have been generated by 
the process of the gravitational instability, acting on those initially
small perturbations.

The goal of this paper is to investigate in direct numerical simulations 
and theoretical analysis the validity of the adhesion model, a benchmark
for the non linear development of the gravitational instability.
We will make contact with recent results of Buchert \& Dom\'inguez 
(\cite{Buchert1}) and present a sharpening of their main findings.
The conclusion of this study is that the Burgers equation of the adhesion 
model does not model structure formation quantitatively correctly, but
another transport equation of a similar type does.

We will limit ourselves to a flat, critical 
Universe ($\Omega=1$) where most mass is contained in a dark component. 
In fact, we assume
effectively that all the matter is dark, or behaves as such. There is no 
vacuum energy, or, equivalently, the cosmological constant is zero. 
This has been the favored cosmological model in the recent past 
and, although it is not at the present, it is still of conceptual 
and qualitative interest. We remark that the approximation of a flat
critical Universe will be an accurate one for most of the time
since recombination epoch, also in the presently  favored models.

We will further make the assumption of stratified perturbations. This
is surely not realistic, but, as we will see, it allows for a numerical scheme
of incomparable speed and accuracy, describing the full process of
structure formation, and is therefore a useful testing ground.

The paper is organized as follows: in Section \ref{s:back} we present the 
general background. In Section \ref{s:q} we recall the  derivation of the 
Quintic model (Aurell \& Fanelli \cite{Qmodel}), 
that allows to study the evolution of a stratified perturbation in a 
Einstein-de Sitter, three dimensional expanding Universe. 
Section \ref{s:ns} is devoted to the discussion of the numerical
implementation. In Section \ref{s:scal} we present the result of our 
numerical study: scaling laws are displayed and an heuristic explanation
is provided. In Section \ref{s:gr} we derive an equation of transport
that is consistent with the numerical findings. In the final Section
\ref{s:con} we sum up and discuss our results.

\section{General Background}
\label{s:back}

The linear regime of structure formation was studied by 
Lifshitz \cite{Lifshitz} and is described in several classical monographs
(Peebles \cite{Peebles1980}, Weinberg \cite{Weinberg1972}). For the non 
linear regime, there is a long tradition of considering various simplified 
models, from the Zeldovich pancake (Shandarin \& Zeldovich \cite{Zeldovich}) 
and the adhesion model ( Gurbatov, Saichev \&  Shandarin \cite{Gurbatov}), 
to full-scale $N$-body simulations, with the accompanying approximations of 
numerical nature. We start with a short review of the basic setup and 
equations.

Collision-less dark matter is described by the kinetic Vlasov-Poisson 
equations, in an expanding three dimensional Universe. This level of 
approximations contains the assumption that tensor degrees (gravitational
waves) can be neglected, and that the matter motion is quasi-Newtonian. 
We assume an 
inertial reference frame, and label the position with ${\bf r}$.
Then  it is customary to introduce the comoving 
coordinate ${\bf x}$ by performing the following transformation
(Peebles \cite{Peebles1980}):
\begin{equation}
{\bf r} = a(t) {\bf x}~,
\end{equation}
where the scale factor $a(t)$ is function of 
proper world time. For the critical Friedman Universe:
\begin{equation}
\label{scale}
a=\left(\frac{t}{t_0}\right)^{2/3}~,
\end{equation}
and $t_0^{-2}=6 \pi G \rho(t_0)$, where $\rho(t_0)$ is the homogeneous 
density at time $t_0$ 
(Peebles \cite{Peebles1980}, Weinberg \cite{Weinberg1972}) and $G$ is the 
gravitational constant. The  Vlasov-Poisson equations read:
\begin{equation}
\label{eq:system-JVP2}
\left\{
\begin{array}{l}
\displaystyle{\partial_t f+\frac{\bf p}{m a^2}\cdot
{\bf \nabla_x} f -{\bf \nabla_x}\psi\cdot
{\bf \nabla_p} f = 0}
\\
\\
\displaystyle{\nabla^2\psi=4 \pi G a^2 \left( \rho - \rho_b \right)}~,  
\end{array}
\right.
\end{equation}
where ${\bf p}$ is the variable conjugated to ${\bf x}$; 
$f({\bf x},{\bf p},t)$ is the distribution
function in the six dimensional phase space $({\bf x},{\bf p})$;  
$\psi$ is the gravitational potential and $\rho_b$ is 
the mean mass density distribution.
The particle density $\rho({\bf x},t)$ and velocities ${\bf u}({\bf x},t)$ 
are given, in term of $f({\bf x},{\bf p},t)$, as:
\begin{equation}
\label{rho0}
\rho({\bf x},t)=\frac{m}{a^3} \int f({\bf x}, {\bf p},t) d{\bf p}~,
\end{equation}
\begin{equation}
\label{u}
\rho({\bf x},t) {\bf u}({\bf x},t) =
\frac{1}{a^4} \int {\bf p} f({\bf x},{\bf p},t) d{\bf p}.
\end{equation}

It is well known that (\ref{eq:system-JVP2}) admits  special solutions of 
the form (Vergassola, Dubrulle, Frisch \& Noullez~\cite{Vergassola}):
\begin{equation}
\label{singlespeed}
f({\bf x},{\bf p},t) = \frac{a^3 \rho({\bf x},t)}{m} \delta^d ({\bf p}-m a 
{\bf u}({\bf x},t))~,
\end{equation}
where $d$ is the dimension of space and $\delta^d(.)$ the $d$-dimensional 
delta function. We will refer to this class as single-speed solutions, 
because to each given (${\bf x},t$) corresponds a well defined velocity
${\bf u}$. Assuming (\ref{singlespeed}), after some manipulation, it follows
from equations (\ref{rho0}) and (\ref{u}):
\begin{equation}
\label{s:system}
\left\{
\begin{array}{l}
\displaystyle{\partial_t\rho+3\frac{\dot{a}}{a}\rho +
\frac{1}{a}{\bf \nabla} \cdot  ({\rho} {\bf u})=0}
\\
\\
\displaystyle{\partial_t {\bf u}+
\frac{\dot{a}}{a}{\bf u}+
\frac{1}{a}({\bf u} \cdot {\bf \nabla}) 
{\bf u} = {\bf g}}
\\
\\ 
\displaystyle{{\bf \nabla} \cdot {\bf g}=-4 \pi G a (\rho-\rho_b)}~,  
\end{array}
\right.
\end{equation}
where we have introduced ${\bf g} = - {\bf \nabla} \psi/a$ such that 
${\bf \nabla} \times {\bf g} =0$. It should be stressed that system
(\ref{s:system}), is valid as long as the distribution function 
$f({\bf x},{\bf p})$ is in the form (\ref{singlespeed}). 
Beyond the time of caustic formation, when the fast particles cross the slow 
ones, the solution become multi-stream. 
Hence, the pressure-less and dissipation-less hydrodynamical equations 
(\ref{s:system}) are incomplete. We will in this paper extend system 
(\ref{s:system}) beyond caustic formation for stratified flows, i.e. when 
velocity has one component only, and varies with respect to this direction. 

Let us focus now on time before the first particle crossing. Then we can
further make the assumption of parallelism:
the peculiar velocity is  a potential field,  which remains 
parallel to the gravitational peculiar acceleration field 
(Buchert~T., Dom\'inguez~A. \& Perez-M\'ercader \cite{Buchert2}, Peebles 
\cite{Peebles1980}, Vergassola, Dubrulle, Frisch \& Noullez 
\cite{Vergassola}):
\begin{equation}
\label{parallel}
{\bf g}=F(t){\bf u}~,
\end{equation}
where $F(t)$ is a positive, time dependent, proportionality coefficient.  
Relation (\ref{parallel}) is well justified in the linear, as well in the 
weakly non linear regimes and allows to treat analytically the problem.
From the linear theory it follows (Buchert~T., Dom\'inguez~A. \& 
Perez-M\'ercader \cite{Buchert2}, Peebles \cite{Peebles1980}, 
Vergassola, Dubrulle, Frisch \& Noullez \cite{Vergassola}):
\begin{equation}
\label{parall}
F(t)=4 \pi G \rho_b b/\dot{b}~,
\end{equation}
where $b$ represents the growing mode of the density field in the linear 
regime. Hence, defining the new velocity field 
${\bf v}={\bf u}/(a \dot{b})$, the system (\ref{s:system}) reduces
to:
\begin{equation}
\label{burgers}
\partial_b {\bf v} + ({\bf v} \cdot {\bf \nabla}) {\bf v}=0~,
\end{equation}
which is the multidimensional Burgers equation (Vergassola~M., Dubrulle~B., 
Frisch~U. \& Noullez~A. \cite{Vergassola}). 
The inviscid Burgers equation describes the free motion of fluid particles 
subject to zero forcing and is equivalent to the famous Zeldovich 
approximation (Shandarin \& Zeldovich \cite{Zeldovich}). Again, it is 
worth recalling that the picture is correct as long as the solution stays 
single-stream. After caustic formation, it has been proposed that the 
resulting change on the gravitational force can be modeled by an 
effective diffusive term
(adhesion model (Gurbatov~S.N., Saichev~A.I. \&  Shandarin~S.F. 
\cite{Gurbatov})). This should represent the effect of the 
gravitational sticking not captured by the Zeldovich approximation.
Mathematically, this means introducing 
a  term of the form  $\mu \nabla^2 {\bf v}$, in the right hand side of the 
equation (\ref{burgers}). In order for the diffusion term to have a 
smoothening effect 
only in those regions where the particles crossing takes place, the viscosity 
$\mu$ should be small. The 
adhesion model reads:
\begin{equation}
\label{s:adhesion}
\left\{
\begin{array}{l}
\displaystyle{\partial_b {\bf v} + ({\bf v} \cdot {\bf \nabla}) {\bf v}  
= \mu \nabla^2 {\bf v} }
\\
\\
\displaystyle{{\bf v}= - {\bf \nabla} \tilde{\psi}}
\\
\\ 
\displaystyle{\partial_b \rho + {\bf \nabla} \cdot (\rho {\bf v})=0 }~.  
\end{array}
\right.
\end{equation}
where $\tilde{\psi} = \psi/(\dot{b} F(t) )$. 
The limit when viscosity $\mu$ tends to zero is often taken. We remark, that,
as is well known this is not equivalent to setting $\mu$ to zero
from the outset, but is instead a regularization of (\ref{burgers}), 
equivalent to the Lax entropy condition. 

Although numerical experiments suggest qualitative agreement, no theory 
is to our knowledge presently available that quantifies the exact 
relationship between (\ref{eq:system-JVP2}) and (\ref{s:adhesion}), after 
caustic formation.

The hypothesis has been put forward that (\ref{s:adhesion}) is an 
asymptotic description of (\ref{eq:system-JVP2}), for particular classes of
initial conditions (Starobinsky \cite{starobinsky}). 
This statement relies on the assumption 
that, because of the expansion, the thickness of the individual pancakes 
grows more slowly than their typical separation. We will refer to this 
picture as to  Starobinsky conjecture. 
Repeating the statement in the introduction, the aim of this paper is 
to investigate the asymptotic relation between the Vlasov approach 
(\ref{eq:system-JVP2}) and the adhesion model 
(\ref{s:adhesion}), focusing our attention on the stratified
dynamics. To jump also again to the conclusion, we will derive transport 
equation similar to those of Buchert\& Dom\'inguez (\cite{Buchert1}), but 
not quite the same.

\section{Quintic model}
\label{s:q}

We will study stratified perturbation in a critical Universe, in a particle 
representation. We recall that a non-linear change of time allows us 
to transform the equations of motion to ordinary differential equations 
with constant coefficients, and which can therefore be integrated by a fast 
event-driven scheme. We will refer to this dynamics as to the Quintic model
(Aurell \& Fanelli \cite{Qmodel}), for reason that will become clear.

The Newtonian equations of motion for $N$ particles follow from a 
Lagrangian
\begin{equation}
\label{background_lagrangian}
{\mathcal{L}} = \sum_i \frac{1}{2} m_i \dot{{r_i}}^2 - 
m_i \phi ({r_i},t)
\end{equation}
where $\nabla_r^2\phi = 4\pi G\rho$. 
In the point particle picture the density profile is: 
\begin{equation}
\label{rho}
\rho(x_i,t)=\sum_{x_j} m_j a^{-3} \delta(x_i-x_j)~,
\end{equation}
where $x_i$ is the comoving coordinate of the $i-th$ particles, in the 
direction of which the density and velocities varies. 

Expressing (\ref{background_lagrangian}) as function of the proper 
coordinate, $x_i$, and assuming (\ref{rho}), the equation of motion of the 
$i$-th  particle reads: 
\begin{equation}
\label{euler1}
\frac{d^2 x_i}{dt^2} + 2 \frac{\dot{a}}{a} \frac{d x_i}{dt}
-4 \pi G  \rho_b(t) x_i = a^{-3} E_{grav}(x_i,t)~,
\end{equation}
where
\begin{equation}
\label{selfgrav}
E_{grav} (x_i,t) = - 2 \pi G  \sum_j m_j \hbox{sign} (x_i-x_j)~.
\end{equation}
From the equation of continuity:
\begin{equation}
\label{cont}
\rho_b(t) = \rho_0 a(t)^{-3}~,
\end{equation}
and, by performing a suitable non linear transformation of the time 
variable it is possible to concentrate all the time dependence in the term
$2 \frac{\dot{a}}{a} \frac{d x_i}{dt}$. The choice is:
\begin{equation}
\label{Rouet-time}
dt = a^{3/2} d \tau~, 
\end{equation}
where $\tau$ has dimension of time ( Rouet, Feix \& Navet \cite{Rouet1990}, 
Rouet et al. \cite{Rouet1991}). 
The equation (\ref{euler1}) is thus transformed into:
\begin{equation}
\label{euler2}
\frac{d^2 x_i}{d \tau^2} + \frac{\dot{a}\sqrt{a}}{2} \frac{d x_i}{d \tau}- 4 \pi G \rho_0  x_i = E_{grav} (x_i,\tau)~.
\end{equation}
In a flat Einstein de Sitter model the scale factor $a(t)$ grows with 
time as a power-law (see eq. (\ref{scale}))
and therefore eq. (\ref{euler2}) takes the form:
\begin{equation}
\label{euler3-Q}
\frac{d^2 x_i}{d \tau^2} +  \frac{1}{3t_0} \frac{d x_i}{d \tau}
-  \frac{2}{3t_0^2} x =  E_{grav} (x_i,\tau) \qquad\hbox{Q model}~. 
\end{equation}

The possibility of making this coordinate change is indeed the technical 
reason why we work with a critical Universe.
Equation (\ref{euler3-Q}) is the model we refer to as to the 
{\it Quintic (Q) model}.
We stress that the quintic model is nothing but a particle picture of 
the full self-gravitating dynamics for the class of stratified perturbations.
The continuum ($N \rightarrow \infty$) limit of (\ref{euler3-Q}) 
is just (\ref{eq:system-JVP2}), with different time coordinate.
The interest of this formulation is that, as for the 
classical static self-gravitating systems in one dimension,
$E_{grav}$ is a Lagrangian invariant, proportional to the net mass 
difference to the right and to the left of a  given particle at a given 
time. Hence, the evolution of the system is recovered by identifying the 
time and location of the particles crossings, and connecting 
analytical solutions between such events (Noullez, Fanelli \& Aurell 
\cite{al}).

\section{Numerical scheme}
\label{s:ns}

The equation of motion of each particle, in the Q model, is specified 
by equation (\ref{euler3-Q}). In between successive crossings 
($\tau \in [\tau^n, \tau^{n+1}]$) the right hand side is a 
constant and, therefore, (\ref{euler3-Q}) admits an explicit solution in 
the form (Aurell \& Fanelli \cite{Qmodel}):
\begin{equation}
\label{global}
x_i(\tau) = c^i_1 \exp(\frac{2(\tau-\tau^n)}{3t_0}) + c^i_2 \exp(-\frac{(\tau-\tau^n)}{t_0})- K_i^n~, 
\end{equation}
where $K_i=(3t_0^2 / 2) E_{grav} (x_i,\tau)$, is constant.
The coefficients $c^i_1$ and $c^i_2$ are determined by $x_i^n=x_i(\tau^n)$ 
and $w_i^n=\dot{x_i}(\tau^n)$, i.e. by the states of the particle at the 
time of the last crossing, and read: 
\begin{equation}
\label{s:coeff0}
\left\{
\begin{array}{l}
\displaystyle{c^{i}_1 =  \frac{3}{5} \left[ x_i^n + t_0 w_i^n  - K_i\right]}
\\
\\
\displaystyle{c^{i}_2 = \frac{2}{5} \left[ x_i^n - \frac{2}{3} t_0 w_i^n  - 
K_i \right]  }~.
\end{array}
\right.
\end{equation}

The form of equation (\ref{global}) suggests introducing an 
auxiliary variable $z=\exp( (\tau-\tau^n) / 3t_0 )$. The crossing times 
between neighboring particles (i.e. $i$,$i+1$) can, hence, be computed by 
solving numerically a quintic equation in the form:
\begin{equation}
\label{quint}
f(z) = A_{i,i+1}^n z^5 - B_{i,i+1}^n z^{3} + C_{i,i+1}^n = 0~,
\end{equation}
where:
\begin{equation}
\label{s:coeff}
\left\{
\begin{array}{l}
\displaystyle{A_{i,i+1}^n =  \frac{3}{5} \left[ 
\Delta x_{i}^n + t_0 \Delta w_{i}^n  - (K_{i+1}-K_i) \right]}
\\
\\
\\
\displaystyle{B_{i,i+1}^n =  -(K_{i+1}-K_i) = const}
\\
\\ 
\displaystyle{C_{i,i+1}^n =\frac{2}{5} \left[ 
\Delta x_{i}^n -\frac{2}{3} t_0 \Delta w_{i}^n  - (K_{i+1}-K_i) \right]}~, 
\end{array}
\right.
\end{equation}
and $\Delta x_{i}^n=x_{i+1}^n-x_i^n$,  $\Delta w_{i}^n=w_{i+1}^n-w_i^n$. 
The function $f(z)$ represents the distance between particles $i$ and 
$i+1$, at transformed time $z$.

The event-driven scheme (Noullez, Fanelli \& Aurell \cite{al}) can  be 
adopted to follow the dynamics of the system. The positions of the 
particles are stored in monotonically increasing order.  
A proper time, $\theta_i$, is associated to each  
particle $i$: it refers back to the time the particle last experienced a 
collision. Initially all $\theta_i$ are set to zero. The algorithm 
computes first the crossing time of each particle with the neighbor to 
the right, by solving $N-1$ quintic equations, as in (\ref{quint}) 
(see above).  The results are then stored in an array, which is sorted 
on a heap. Once the heap has been built, the minimum collision time, 
$t_{min}$, is  at the root, i.e. at the first position in the heap. The 
particles  involved in the first collision are selected by means of a 
trivial ${\cal O}(1)$ operation. The algorithm lets them evolve up to 
$t_{min}$, according to the equation (\ref{global}). At this point, the 
particles share the same spatial position, and the crossing takes place
(the velocities are exchanged). The successive step is to compute the next 
predicted collision time between $i$ and $i+1$.  The new value  replaces the 
old one, and the heap needs to be rearranged. In addition, as an effect of the 
changes in velocity for particles $i$ and $i+1$ particles, 
the two collisions with their nearest neighbors (respectively $i-1$ and 
$i+2$) need to be re-computed. The heap is then re-arranged 
with at most ${\cal O}(\log(N))$ operations and the 
whole procedure can be repeated for $N_{coll}$ collisions. Therefore the 
complexity of the algorithm is in worst-case  ${\cal O} (N_{coll} \log (N))$
(Noullez, Fanelli \& Aurell \cite{al}).

Indeed, in this particular application, the numerical procedure for 
finding the solution of the quintic equation might converge slowly,
affecting the whole computation time, with a non negligible contribution.
Therefore, in order to achieve the goal of a fast implementation, 
special care has to be devoted to the analysis of (\ref{quint}). 
Recalling the definition of $f(z)$, we look for the smallest real 
value $\bar{z}>1$, such that $f(\bar{z})=0$. 
As a first trivial observation, we notice that, by definition, $z$ 
is  larger than one, and the inter-particle distance  $f(z=1)$ is 
non-negative. In addition,  $B_{i,i+1}^n$ is a positive 
constant, independent of $i$.

\begin{figure}[ht!]
\begin{center}
\includegraphics[width=8cm]{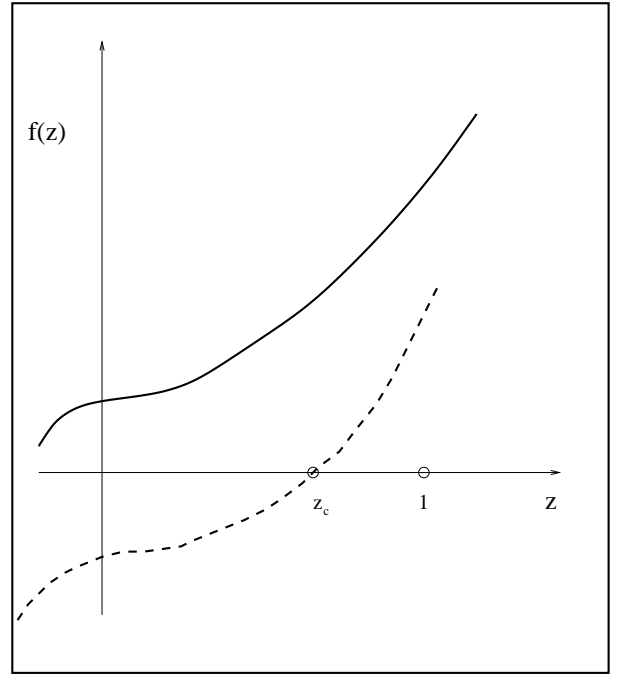}
\caption{\em \label{caso1} The function $f(z)$  represented 
versus  $z$, for  $A_{i,i+1}^n$  positive. The solid line refers 
to the case when  $C_{i,i+1}^n$ is positive while the dashed line to  
$C_{i,i+1}^n$ negative. Note that no intersection with the horizontal axis 
is allowed for $z>1$ (since $f(1)>0$, by definition).}
\end{center}
\end{figure}

Two scenarios are therefore possible: if the coefficient $A_{i,i+1}^n$ is 
positive, no crossing is allowed. As it is sketched in
Fig. \ref{caso1}, the only real root of eq.  (\ref{quint}) lies 
in the interval  $[0,1]$, and therefore has to be rejected. 
This means that these two particles will never collide if left to 
themselves: the expansion is too strong to be overcome by their
mutual gravitational attraction.

\begin{figure}[ht!]
\begin{center}
\includegraphics[width=8cm]{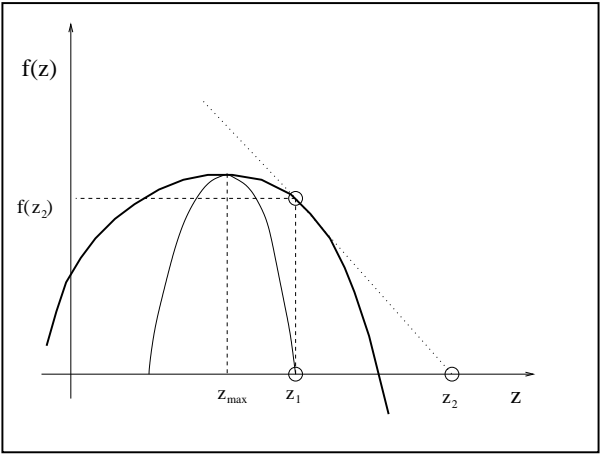}
\caption{\em \label{caso2} The function $f(z)$  represented 
versus  $z$, for $A_{i,i+1}^n$ negative (thick solid line). The 
thin line is the quadratic approximation around $z=z_{max}$. Here $f(z_1)$ 
is positive and therefore, the root $\bar{z}$ is bounded between 
$[z_1, z_2]$. Here $z_2$ is the intersection with the $z$ axis of the
tangent to $f(z)$ in $z_1$ (dotted line).}
\end{center}
\end{figure}

On the other hand, if the coefficient $A_{i,i+1}^n$ is negative, more 
care is required. The problem is to bound the root in a reasonable 
interval in order to assure a fast convergence of a numeric procedure. 
First, we observe that there is a local maximum for $z>0$. The coordinate 
$z_{max}$ is easily computed and $f(z_{max})$ is evaluated;  
$f(z_{max})$ is positive, since it is by definition larger of $f(1)>0$. 
Since the function $f(z)$ goes to $-\infty$ as $z \rightarrow \infty$, 
there should be an intersection ($\bar{z}>1$ s.t. $f(\bar{z})=0$) with the 
horizontal axis, see Fig \ref{caso2}. 
\begin{figure}[ht!]
\begin{center}
\includegraphics[width=8cm]{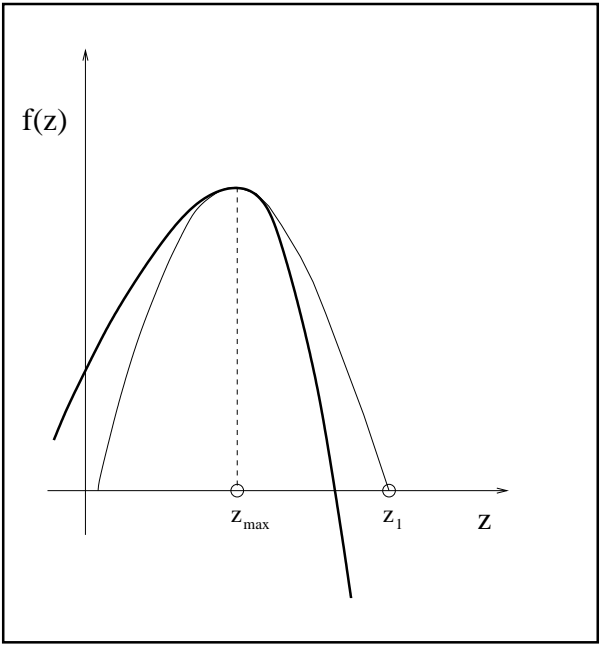}
\caption{\em \label{caso3}The function $f(z)$  represented 
versus  $z$, for  $A_{i,i+1}^n$  negative (thick solid line). 
The thin solid line represents the quadratic approximation around $z=z_{max}$.
Here $f(z_1)$ is negative and, therefore, the root $\bar{z}$ is bounded 
between $[z_{max}, z_1]$.}
\end{center}
\end{figure}
The following procedure is adopted. First we introduce $h(z)$ 
(thin solid line in Figs. \ref{caso3}, \ref{caso2}), that is the 
quadratic approximation of $f(z)$ around $z_{max}$, defined by:

\begin{equation}
h(z)=f(z_{max}) +\frac{1}{2} f^{''}(z_{max}) (z-z_{max})^2~.
\end{equation}

Then we compute the intersection $z_1$ of $h(z)$ with the horizontal axis:

\begin{equation}
z_1=z_{max} + \left ( \frac{f(z_{max}}{3 B_{i,i+1}^n}\right)
\sqrt{ \frac{-5 A_{i,i+1}^n} {3 B_{i,i+1}^n} }~.
\end{equation}

Two different cases have to be considered 
(respectively, Figs. \ref{caso3},\ref{caso2}).
If $f(z_1)$ is positive, we 
take the tangent to $f(z)$ in $z_1$, and compute its intersection, $z_2$, 
with the $z$ axis (Fig. \ref{caso2}):
\begin{equation}
z_2=z_1-\frac{f(z_1)}{5 A_{i,i+1}^n z_1^4 +3 B_{i,i+1}^n z_1^2}~.
\end{equation}

By construction $z_2>\bar{z}$ and, therefore, $\bar{z} \in [z_1, z_2]$.
If, on the contrary, $f(z_1)$ is negative, then $\bar{z} \in [z_{max}, z_1]$,
see Fig. \ref{caso3}.

In both cases we have confined the root in a narrow interval: there, 
$f(z)$ is a monotonic decreasing function. Therefore, we can apply 
a combination of bisection and Newton-Rapson method (Press \cite{num}). This 
hybrid algorithm assures a stable and fast convergence to the solution.
In the present implementation we assumed a tolerance error of $10^{-13}$.

\section{Asymptotic scaling laws: numerical results and heuristic interpretation}
\label{s:scal}

We simulate the dynamics of the Q model, by using  the numerical scheme 
discussed above.  We consider a system of $N$ particles of mass 
$m=1/N$, confined in a box of size $L$. Reflecting boundary 
conditions are assumed, which is equivalent to consider periodic 
perturbation of size $2L$, with reflexion symmetry (Aurell \& Fanelli 
\cite{Qmodel}). 
We choose units such that $4 \pi G$ is equal to one. Time is measured by
the a-dimensional quantity $(t/t_0)$. The  unit of length is
the spatial interval in which the particles are initially distributed 
(i.e. $L=1$), and thus the initial density $\rho_0$ is set to one.

In particular, we are interested in the late time evolution of the system 
to better understand the validity of the Starobinsky conjecture, 
the aim of this analysis being  to provide a quantitative test of the 
reliability of the adhesion model, as an effective description of the
mechanism of large scale structure formation in the Universe. 

With this in mind, we consider the evolution of an individual cluster and 
investigate the process of gravitational collapse. We measure   
the progressive contraction of the inner region of the agglomeration,  
compared to the overall expansion. This effect can be computed  
by the ratio between the width of region, $\Delta x$, that 
contains half of the whole mass of the system, centered around the position 
of maximum density, and the size of the perturbation  $L$ 
\footnote{In comoving coordinate $L$ is a constant, which we 
have set to one. }. This quantity is then plotted, as function of the 
rescaled cosmological time $t/t_0$. 

Simulations are performed for two different classes of initial conditions. 
In both cases, we make the non restrictive choice of considering a 
perturbation centered around our center of reference.

\begin{figure}[ht!]
\begin{center}
\includegraphics[width=8cm]{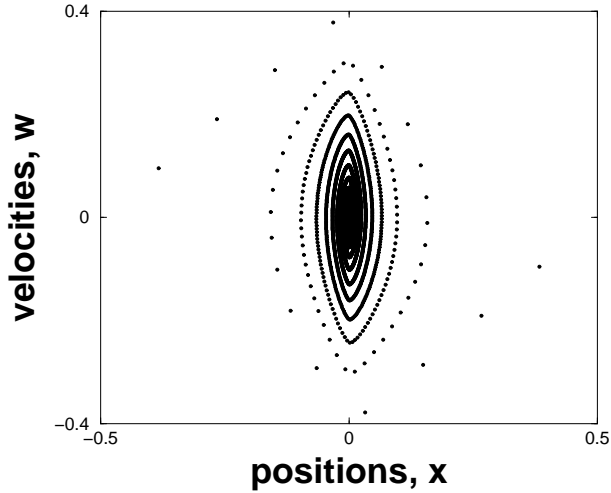}
\caption{\em \label{phase} Velocity field versus positions, starting from a 
single speed initial condition (sinus wave). Here $N=4096$ and 
$t/t_0 = 6.1249064\times 10^4$. Reflecting boundaries are assumed.
 Positions and 
velocities are in arbitrary units.}
\end{center}
\end{figure}

\begin{figure}[ht!]
\begin{center}
\includegraphics[width=8cm]{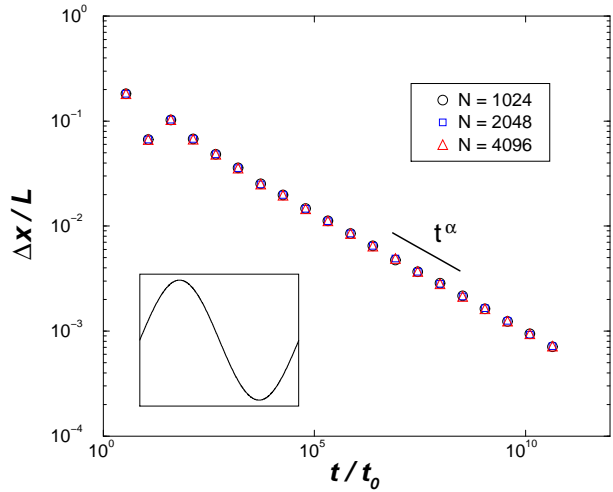}
\caption{\em \label{deltax} $\Delta x/L$ is plotted versus  $t/t_0$. 
Different symbols refer to different values of $N$ (see the legend).
A clear power-law behavior is displayed. The best numerical fit gives 
exponent $\alpha=-0.22$. The small inset represent the initial condition 
in the phase space $(x,w)$.}
\end{center}
\end{figure}

\begin{figure}[ht!]
\begin{center}
\includegraphics[width=8cm]{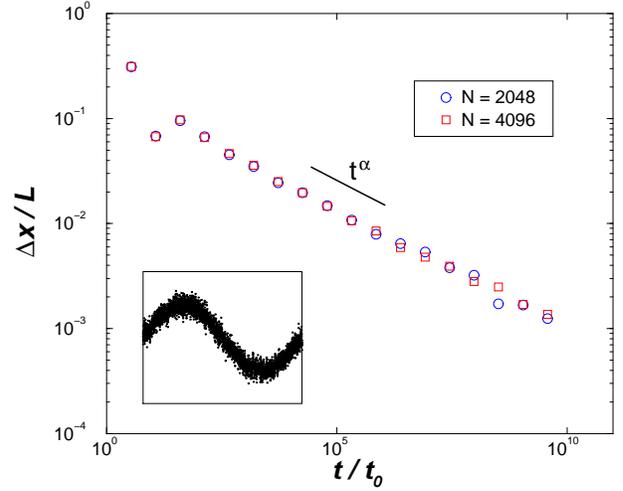}
\caption{\em \label{deltaxnoise} $\Delta x/L$ is plotted versus  $t/t_0$. 
Here $w_0$ represents the amplitude of the initial smooth sinus to which the 
white noise, $s_{wn}$, is superposed (i.e., $w= w_0 sin(x)+s_{wn}$, 
see small inset). Different symbols refer to different values of $N$ 
(see the legend). The scaling is consistent with the one derived from 
Fig. \ref{deltax} (here represented by the thick solid line).}
\end{center}
\end{figure}

First, the initial velocity is assumed to be a smooth  function of position. 
As clearly expected, the system develops spiral in the phase space
like the one displayed in Fig. \ref{phase}. 
In Fig. \ref{deltax} we plot 
$\Delta x/L$,   versus  $t/t_0$. The experiments are performed for 
different values of $N$. 
A clear power-law behavior:
\begin{equation}
\label{scaling_x}
\frac{\Delta x}{L} = \left( \frac{t}{t_0} \right) ^\alpha
\end{equation}
is always displayed, regardless of the number of particles simulated.
This implies that the result is not affected by the discreteness of the 
representation, and can be assumed to hold in the continuum limit.
The scaling  is present over several decades and the best numerical fit 
gives the value $\alpha=-0.22$. 
Then, a source of noise is introduced in the initial condition: 
as shown in the small inset of Fig. \ref{deltaxnoise} a white noise signal 
is generated and superposed to a sinus wave of amplitude $w_0$. 
The thickness of the dense region is studied, following the line of the 
preceding discussion. Results are reported in the main plot of Fig. 
\ref{deltaxnoise}, for a single realization. They show  complete 
agreement with (\ref{deltax}) . 
It is worth stressing that these results are not 
sensitive to the choice of measuring the interval that contains $N/2$
particles. Any other finite fraction leads to the same 
conclusions.

\begin{figure}[ht!]
\begin{center}
\includegraphics[width=8cm]{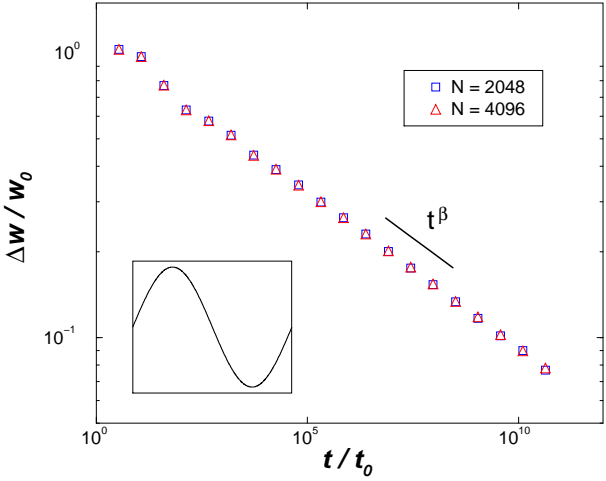}
\caption{\em \label{deltav} $\Delta w/w_0$ is plotted versus  $t/t_0$.
Here $w_0$ represents the amplitude of the initial velocity perturbation
(i.e., $w= w_0 \sin(x)$, see small inset). Different symbols refer to 
different values of $N$ (see the legend). A clear power-law behavior is 
displayed. The best numerical fit gives exponent $\beta=-0.11$}
\end{center}
\end{figure}

\begin{figure}[ht!]
\begin{center}
\includegraphics[width=8cm]{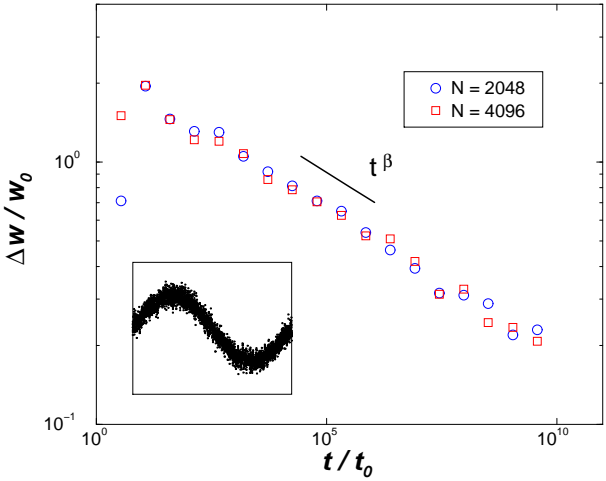}
\caption{\em \label{deltavnoise} $\Delta w/w_0$ is plotted versus  $t/t_0$.
Here $w_0$ represents the amplitude of the initial smooth sinus to which the 
white noise, $s_{wn}$, is superposed (i.e., $w= w_0 \sin(x)+s_{wn}$, 
see small inset). 
Different symbols refer to different values of $N$ (see the legend). 
The scaling is consistent with the one derived from Fig. \ref{deltav}
(here represented by the thick solid line).}  
\end{center}
\end{figure}

In order to provide a full picture, we performed similar analysis for the 
velocities distribution. Consider  $\Delta w$,  such that
$N/2$ particles have velocities in the interval 
$[\Delta w/2 , -\Delta w/2]$.  The ratio $\Delta w / w_0 $ is plotted vs.
$t/t_0$, where $w_0$ represents the amplitude of the initial smooth wave
(see captions of Figs. \ref{deltav} and \ref{deltavnoise}). 
Again, and for both the initial conditions considered here,
a  power-law behavior: 
\begin{equation}
\label{scaling_v}
\frac{\Delta w}{w_{0}} = \left( \frac{t}{t_0} \right) ^\beta 
\end{equation}
is clearly displayed (Figs. \ref{deltav}, \ref{deltavnoise}). 
Here the best numerical fit gives $\beta=-0.11$. 


Assuming the occurrence of power-law behaviors, there is a simple, 
heuristic, explanation for deriving the correct value of the exponents, 
in agreement with the numerical findings. 
As previously stated, the Quintic model is equivalent to the Vlasov-Poisson 
set of equations in the continuous limit (Aurell \& Fanelli \cite{Qmodel}). 
Such a system is conservative and therefore the volumes in the phase space 
are conserved (Liouville theorem). Hence:
\begin{equation}
\label{liouville0}
\Delta x \Delta p = const~,
\end{equation}   
where $\Delta x$ and $\Delta p$ represent, respectively, any space and 
momentum intervals in the comoving reference frame. 

Recalling that $w=d x/d \tau$ and that $a(t)=(t/t_0)^{2/3}$, 
eq. (\ref{liouville0}) is transformed into:
\begin{equation}
\label{liouville}
\Delta x \Delta w = \left( \frac{t}{t_0} \right) ^{-1/3}~.
\end{equation} 

Assuming that $\Delta x$ and  $\Delta w$ scales in time as power-laws, 
respectively with exponents $\alpha$ and $\beta$ it follows that:
\begin{equation}
\label{rel1}
\alpha+\beta=-\frac{1}{3}~.
\end{equation} 

A second relation is needed to close the system. 
Let us consider the Hamiltonian $H = K + V$ which describes the
Quintic model, except the friction term. $K$ is a normal kinetic term, 
quadratic in the velocities, and $V$ is made of two terms: the interaction 
and the background potential. The following relation holds:
\begin{equation}
\label{DH}
\frac{dH}{dt} = - \frac{2}{3 t_0} K~.
\end{equation}

Assume that the global virialization has occurred.
Then $<K> \sim <V>$, where  $<K>$ and $<V>$
represent the time average of the kinetic and potential energies.
$K$ is quadratic in velocities while $V$ has one term quadratic (the
background), and one linear (the interaction). If we have a mass 
concentration at the origin, the quadratic can be
ignored compared to the linear.

Hence, neglecting the time averages, we have $velocities^2 \sim distances$. 
If $\Delta x \sim (t/t_0)^\alpha$, therefore 
$\Delta w \sim (t/t_0)^{\alpha/2}$, which implies:
\begin{equation}
\label{rel1}
\beta=\frac{\alpha}{2}~,
\end{equation} 
and therefore $\alpha=-2/9$ and $\beta=-1/9$, in agreement with 
the numerics. From eq. (\ref{DH}), the dissipative time is:
\begin{equation}
T_{diss} = H/(dH/dt) \sim constant~.
\end{equation}

The virialization time of the mass agglomeration, $T_{vir}$  is about 
the time it takes one particle to traverse the mass, that is 
\begin{equation}
T_{vir} = (distance)/(velocity) \sim \left( \frac{t}{t_0} \right)^{\beta}~. 
\end{equation}

Hence, the virialization time becomes quickly smaller than the
dissipative time, and the argument is self-consistent.

\section{Transport equation }
\label{s:gr}

As already observed,  relation (\ref{singlespeed}) holds as long as
the solution stays single stream. Hence the hydrodynamical picture
(\ref{s:system}) is applicable just before the time of caustic 
formation, when the first particles crossing takes place. In order to 
extend the analysis beyond that time we have to consider the more 
general solution to (\ref{eq:system-JVP2}):
\begin{equation}
\label{multispeed}
f({\bf x},{\bf p},t) = \frac{a^3 \rho({\bf x},t)}{m} f_0({\bf x},{\bf p},t)~.
\end{equation}
where $f_0({\bf x},{\bf p},t)$ is the velocity profile.

Assuming (\ref{multispeed}) and performing the same analysis as described in 
Section \ref{s:int}, the following system is derived from  
Vlasov-Poisson equations (\ref{eq:system-JVP2}):
\begin{equation}
\label{s:system1}
\left\{
\begin{array}{l}
\displaystyle{\partial_t\rho+3\frac{\dot{a}}{a}\rho +
\frac{1}{a}{\bf \nabla} \cdot  \left( {\rho} {\bf \overline{u}} \right) =0}
\\
\\
\displaystyle{\partial_t {\bf \overline{u}}+
\frac{\dot{a}}{a}{\bf \overline{u}}+
\frac{1}{a} \left( {\bf \overline{u}} \cdot {\bf \nabla} \right) 
{\bf \overline{u}} = {\bf g} - \frac{1}{a \rho} {\bf \nabla} 
\left[ \rho ( \overline{{\bf  u}^2} - {\bf \overline{u}}^2) \right]}
\\
\\ 
\displaystyle{{\bf \nabla} \cdot {\bf g}=-4 \pi G a (\rho-\rho_b)}~,  
\end{array}
\right.
\end{equation}
where $\overline{[~\cdot~]}$ represents the {\it mean}, after averaging over
velocity space. Comparing with (\ref{s:system}), we see that  
in the second equation of (\ref{s:system1}) an extra-term appears, that takes 
into account the effect of the dispersion in velocities, in the region where 
the multi-values solution is developed. When the flow is single-stream, the 
variance is zero and the system (\ref{s:system1}) is  identical to 
(\ref{s:system}) (${\bf \overline{u}} \equiv {\bf u}$). 
This derivation was already discussed in (Buchert \& Dom\'inguez 
\cite{Buchert1}) 
and represents the starting point of our analysis. 
We will limit ourselves to the case of stratified perturbations, and
derive an equation for the  transport of the velocity 
flow, that  agrees with the numerical analysis performed at the level of the 
Q model (i.e. discrete Vlasov). In this respect, we are mainly concerned by 
the progressive collapse of the inner bulk (see Figs 
\ref{deltax},\ref{deltaxnoise}), a key feature that needs to be considered 
explicitly in our analysis. The question that naturally arises is whether
or not the transport is well represented by Burgers' equation.
Here on, we will indicate the mean velocity simply with ${\bf u}$. 

Let us consider the dynamics of the Q model. It can be shown numerically 
that,  for all classes of initial conditions here considered, far beyond the 
time of first crossing, the multi-stream velocity profile $f_0(w)$ is 
well approximated by a Gaussian:
\begin{equation}
\label{gaussian}
f_0(w) \propto \exp(-\frac{\left(w-\overline{w}\right)^2}{T(x)})~,
\end{equation}
where the {\it temperature }, $T(x)$, is smaller in the bulk 
(more narrow Gaussian) than in the surrounding regions (larger profile). 
Equation (\ref{gaussian}) is also, of course, the natural choice in a system 
close to equilibrium, where $\rho$,$w$ and $T$ vary considerably over
a cluster, but little over distances of the order of inter-particle
spacing. 

\begin{figure}[ht!]
\begin{center}
\includegraphics[width=8cm]{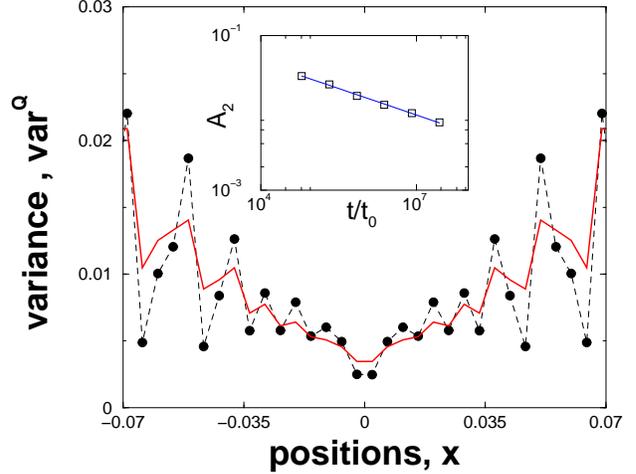}
\caption{\em \label{variance} Main plot: $var^Q(x)$ is plotted versus  $x$.
Here $N=8192$ and $t/t_0 = 6.1249064\times 10^4$. The filled circles 
refers to the numerical experiment. The dashed line is a guide to the 
eye. The solid line is obtained by making use of the fitting function
$A_1 / \rho^{A_2}$, with $A_1,~A_2$, free parameters (refer to text for a 
more detailed description). Here $A_2=0.3$. The small inset represents 
$A_1$ vs $t/t_0$ (squares), obtained keeping $A_2=0.3$ fixed. 
The solid line is a power-law fit that gives exponent 
$\xi = -0.22 \sim -2/9$.}
\end{center}
\end{figure}

Our ansatz for $T(x)$ is the following:
\begin{equation}
\label{ansatz}
T(x)=K \frac{(t/t_0)^\xi}{\rho^\gamma}~,
\end{equation}
where $K$ is a constant and $\gamma$ is a real number belonging to the 
interval $[0,1]$. 
This choice reproduces qualitatively the shrinking of the velocity 
profile, observed in comoving coordinates in correspondence of the denser 
inner regions.  
On a more quantitative level, numerical checks have been performed to 
support the validity of (\ref{ansatz}), showing in all cases,  a 
satisfactory agreement. In particular, from (\ref{gaussian}) it follows that
the variance, $var^Q(x)$, is given by:
\begin{equation}
\label{varian}
var^Q(x)=\overline{w^2}-\overline{w}^2 =\frac{K}{2}
\frac{(t/t_0)^\xi}{\rho^\gamma}~,
\end{equation}
where the label $Q$ indicates that we work in Quintic variables.
In the main plot of figure (\ref{variance}) $var^Q(x)$ is plotted 
versus $x$. The filled circles refers to the results of the  
numerical experiments, while the solid line is a two parameters fitting,  
based on (\ref{ansatz}). (i.e. $var^Q(x) = A_1/\rho^{A_2}$, where now 
$\rho$ is the histogram of positions and $A_1,A_2$ free parameters). 
In Fig. \ref{variance},~$A_2=0.3$: the same value, within the errors 
($\sim 0.1$), is found for different initial realizations and/or different 
times. 
Larger deviations from (\ref{varian}) are observed  outside 
the core, where the density reduces drastically and the multi-stream flow 
approaches the single stream limit. The squares in the small inset 
represent, in double logarithmic scale, the amplitude $A_1$ resulting 
from the fit (keeping now $A_2=0.3$) , for different times. 
The power-law scaling is clear and the exponent is $\xi=-0.22 \sim -2/9$. 
We will come back on this important point in the end of the Section.

Now, let us consider the system (\ref{s:system1}). 
Following the discussion in Section \ref{s:int}, we perform 
the change $v \rightarrow  u/a \dot{b}$ and define the new time 
variable $b$. Then the second equation of (\ref{s:system1}) reads:
\begin{equation}
\label{transport}
\partial_b v+
\left(  v \partial_x \right) v 
= \Gamma^B(x, v)~,
\end{equation}
where $\Gamma^B(x,v)$ given by:
\begin{equation}
\begin{array}{c}
\label{extra}
\Gamma^B(x,v) = -\frac{1}{a^2 \dot{b} \rho} \partial_x \left[ \rho~var^B(x) 
\right] =\\
\\
 -\frac{9}{4 a^4 \dot{b}\rho} \partial_x \left[ \rho~var^Q(x) \right]~. 
\end{array}
\end{equation}

Here $B$ indicates that the quantity is expressed in {\it Burgers-like}  
variables, i.e. $(x,v,b)$. In (\ref{extra}) the transformation 
$w \rightarrow \frac{2}{3} a(t) v$ has been considered explicitly. From 
equation (\ref{varian}) we obtain:
\begin{equation}
\label{extra1}
\Gamma^B(x,v) = -\frac{9 K}{8} \frac{(t/t_0)^\xi}{a^4 \dot{b}^2 \rho} 
\partial_x \left[ \frac{1}{\rho^{\gamma-1}} \right]~.
\end{equation}

Inserting the assumption of parallelism (\ref{parall}), the Poisson 
equation (the third of (\ref{s:system1})) reads:
 \begin{equation}
\partial_x \rho = \frac{F(t)}{4 \pi G a} \partial_{x x} u = 
-\rho_b b  \partial_{x x} v~.
\end{equation}

Hence equation (\ref{extra1}) is transformed into:
\begin{equation}
\Gamma^B(x,v) =  -\frac{9 K}{8} \frac{b (t/t_0)^\xi}{a^4 \dot{b}^2}
\frac{\rho_b}{\rho^{\gamma+1}} \partial_{xx} v~. 
\end{equation}

In addition the following relation holds:
\begin{equation}
\frac{1}{\rho^{\gamma+1}} = \frac{1}{\rho_b^{\gamma+1}} 
\left[ \frac{1}{1-b \partial_x v} \right ]^{\gamma+1}~,
\end{equation}
and, thus:
\begin{equation}
\Gamma^B(x,v) =  -\frac{9 K}{8} (\gamma-1) 
\frac{(t/t_0)^\xi}{a^{4-3 \gamma} \dot{b}^2} \frac{1}{\rho_0}
\partial_x \left[ \frac{1}{1-b \partial_x v} \right ]^{\gamma}~, 
\end{equation}
where we made use of equation (\ref{cont}). Collecting together, 
equation (\ref{transport}) reads:
\begin{equation}
\label{transport1}
\partial_b v+
\left(  v \partial_x \right) v 
= \mu(t) \partial_x \left[ \frac{1}{1-b \partial_x v} \right ]^{\gamma}~,
\end{equation}
where $\mu(t)$ is a time dependent coefficient defined by:
\begin{equation}
\mu(t) = -\frac{9 K}{8} (\gamma-1) \frac{(t/t_0)^\xi}{a^{4-3 \gamma}
\dot{b}^2} \frac{1}{\rho_0} =
a_1 \left( \frac{t}{t_0}\right)^{\alpha(\xi,\gamma)}~.
\end{equation}

Equation (\ref{transport1}) is defined in the inner region where the 
ansatz (\ref{ansatz}) applies and where the gradient of $v$ is negative.
We stress that, even though obtained in a different manner,
(\ref{transport1}) belongs to the same family of equations derived by 
( Buchert, Dom\'inguez \& Perez-M\'ercader \cite{Buchert2}), except for a 
slight modification of 
$\mu(t)$ and a different interpretation of $\gamma$. More important, 
using the constraints imposed by the results in section \ref{s:ns}, we 
will provide a precise indication of the values of $\gamma$ and $\xi$.
We therefore proceed  to select one candidate from the whole family 
(\ref{transport1}). 

Consider equation (\ref{transport1}) and apply the
following rescaling for position, velocity and time: 
\begin{equation}
x= \left( \frac{t}{t_0}\right)^{\lambda_1} \tilde{x}~,~~~~~
v= \left( \frac{t}{t_0}\right)^{\lambda_2} \tilde{v}~,~~~~~
db = \left( \frac{t}{t_0}\right)^{\delta} d \tilde{b}~.
\end{equation}

Equation  (\ref{transport1}) is transformed into:
\begin{equation}
\label{transport2}
\begin{array}{c}
\partial_{\tilde{b}} \tilde{v} + \frac{3}{2} \lambda_2  \left( \frac{t}{t_0}\right)^{\delta-2/3} \tilde{v}+ \left( \frac{t}{t_0}\right)^{\lambda_2-\lambda_1-\delta} \tilde{v} \partial_{\tilde{x}} \tilde{v}=\\
\\
a_1 \left( \frac{t}{t_0}\right)^{\alpha(\xi,\gamma)-\lambda_1-\delta-
\lambda_2} \partial_{\tilde{x}} \left[ 
1- \left( \frac{t}{t_0}\right)^{2/3-\lambda_1+\lambda_2}  
\partial_{\tilde{x}} \tilde{v} \right ]^{-\gamma}~.
\end{array}
\end{equation}

By setting $\lambda_1=\alpha/2+2/3$, $\lambda_2=\alpha/2$ and  
$\delta=-2/3$,  equation  (\ref{transport2}) simplifies:
\begin{equation}
\label{transport3}
\partial_{\tilde{b}} \tilde{v} + \frac{3}{2} \lambda_2 \tilde{v}+ 
\tilde{v} \partial_{\tilde{x}} \tilde{v}= a_1 \partial_{\tilde{x}} 
\left[ \frac{1}{1- \partial_{\tilde{x}} \tilde{v}} \right ]^{\gamma}~.
\end{equation}

All the coefficients are now time independent. Hence, equation
(\ref{transport3}) develops shocks of constant width, 
$\Delta \tilde{x}_{shock}$. Transforming back to the old variables, 
this implies:
\begin{equation}
\Delta x_{shock} = \left( \frac{t}{t_0}\right)^{\alpha/2+2/3} \cdot  \Delta 
\tilde{x}_{shock} \propto \left( \frac{t}{t_0}\right)^{\alpha/2+2/3}~.
\end{equation}

In order to provide a full consistent picture, we  require the shock interval 
to shrink in time with the same rate that have been shown to hold 
for the discrete Vlasov equation (Q model). Therefore we have to impose:
\begin{equation}
-\frac{2}{9} = \frac{\alpha}{2}+\frac{2}{3}~,
\end{equation}
that implies the following relation between $\gamma$ and $\xi$:
\begin{equation}
\label{gamma-beta}
\gamma =- \frac{1}{2} \xi + \frac{1}{9}~. 
\end{equation}

Let us now consider the results reported in Fig. \ref{variance}. Assuming 
$\xi=-2/9$ (see analysis above and small inset in Fig. \ref{variance}), 
from equation (\ref{gamma-beta}) one gets $\gamma=2/9$ that is 
in agreement with the result of the numerical fitting. Therefore we 
are led to conclude that asymptotically, the evolution a of multi-stream
flow originated by the one-dimensional Vlasov-Poisson dynamics, is 
mimicked by a transport equation in the form:
\begin{equation}
\label{transport4}
\partial_b v+
v \partial_x v 
= \mu(t) \partial_x \left[ \frac{1}{1-b \partial_x v} \right ]^{2/9}~,
\end{equation}
where $\mu(t) \propto (t/t_0)^{-16/9}$ and $b=(t/t_0)^{2/3}$. 
As we will show in 
the next paragraph, the adhesion model can be recovered from our ansatz 
(\ref{ansatz}) with a special choice of $\gamma$. Nevertheless both
the value of  $\gamma$ and the consequent rate  of compression of 
$\Delta x_{shock}$ are not consistent with the
numerical study of the Q model reported in Section \ref{s:ns}. 
We are therefore led to conclude that the 
Burgers' equation, in the limit of vanishing viscosity, is only valid 
as a qualitative test model and that, on the other hand, the more 
quantitative phenomenology is, instead, captured by the transport 
equation (\ref{transport4}).

\subsection{The adhesion model: limit of validity}

Let us consider the family of equations (\ref{transport1}) derived from the 
ansatz (\ref{ansatz}). Assume $\gamma=-1$: this is the only possible choice  
to recover a Burgers like equation, according to our approach.  
In fact equation (\ref{transport1}) then
reads:
\begin{equation}
\label{transport5}
\partial_b v+
\left(  v \partial_x \right) v 
= b \mu(t) \partial_{xx} v~, 
\end{equation}
where $\mu(t) = a_1 \left( \frac{t}{t_0}\right)^{\xi-4}$. 

There are two major problems with that result. First the choice of 
$\gamma$ is not consistent with the simulations for the Quintic model. 
In fact, if $\gamma<0$, the width of the Gaussian profile (\ref{gaussian}) 
decreases moving to region of lower density. That is the opposite of what 
we observed. Moreover, since $\xi=-0.22 \sim -2/9$ 
(see figure \ref{variance}), 
 $\mu(t)$ decays too fast to agree with the contraction of $\Delta x$
detected with the discrete Vlasov approach 
\footnote{Here $\mu(t) \propto (t/t_0)^{-32/9}$. By a rescaling procedure
(analogous to the one adopted in the previous section) it can be shown that 
$\Delta x_{shock} \propto (t/t_0)^{-32/9}$.}.

Hence, equation (\ref{transport5}), directly derived from 
(\ref{eq:system-JVP2}), with the assumption (\ref{ansatz}) fails in  
reproducing the essence of peculiar aspects of the dynamics, that have been 
shown to hold in the Vlasov like picture. Nevertheless, as already stressed 
by (Buchert, Dom\'inguez \& Perez-M\'ercader \cite{Buchert2}), one 
should note that the viscous like term vanish as $t \rightarrow \infty$ 
and therefore, at least from a qualitative point of view, the Starobinsky 
conjecture is justified (formally replacing $\nu \rightarrow 0$ in 
(\ref{s:adhesion}) with $t \rightarrow \infty$).

\section{Conclusions}
\label{s:con}

In this paper we discussed the problem of structure formation in a 
three-dimensional expanding Universe, focusing on stratified 
perturbations, in a pressure-less medium. This is done by using extensively
the Q model, a Lagrangian representation that was derived in a recent 
paper (Aurell \& Fanelli \cite{Qmodel}). The Q model is valid in the limit 
where Newtonian mechanics applies and it is shown to be equivalent to the 
Vlasov-Poisson equations for $N \rightarrow \infty$. 

In particular we investigated the asymptotic behavior of an initial smooth
perturbation, by measuring the progressive contraction of the inner region, 
compared to the overall expansion. Clear power-law scalings are 
detected and a heuristic explanation is provided. 
Then we derived an asymptotic transport equation for the velocity, 
consistent with these observations.  By means of a combined numerical 
and  analytical procedure, we obtained equation (\ref{transport4}). 
Moreover, we showed that the Burgers equation with vanishing 
viscosity, can be directly derived from the kinetic theory, by assuming the 
ansatz (\ref{ansatz}). Nevertheless, equation (\ref{transport5}) fails in
reproducing the correct asymptotic scaling observed and, therefore, we are 
led to conclude that the adhesion approach is valid only as an approximate 
model of structure formation. In fact, even though Burgers' equation holds 
outside the shocks, the adhesion picture is shown to agree, 
only at a qualitative level, with the correct 
Vlasov description, in the massive cores. 
On the contrary, inside the shocks, the more quantitative 
phenomenology is captured by the transport equation (\ref{transport4}).

\begin{acknowledgements}
We thank G. Kreiss and A. Schenkel for discussions. 
This work was supported by the Swedish Research Council through grants NFR F~650-19981250 (D.F) and NFR I~510-930 (E.A.).
\end{acknowledgements}


\begin{thebibliography}{}



\bibitem[2001]{Qmodel} Aurell, E. \& Fanelli, D. 2001, submitted to Europ. 
Phys. Jour. B, cond-mat/0106444 

\bibitem[1999]{Buchert1} Buchert~T. \& Dom\'inguez~A. 1999,  A\&A ,  335, 395

\bibitem[1999]{Buchert2} Buchert~T., Dom\'inguez~A. \& Perez-M\'ercader 1999,
  A\&A , 349, 343

\bibitem[2001]{japan} Fanelli~D., Aurell~E. \& Noullez~A. 2001, 
                Proceeding of IAU Symposium 208 

\bibitem[1989]{Gurbatov} Gurbatov~S.N., Saichev~A.I. \&  
                Shandarin~S.F. 1989, MNRAS, 236, 385


\bibitem[1947]{Lifshitz} Lifshitz~E. 1947,  J. Phys. USSR, 10, 116


\bibitem[2001]{al}    Noullez~A., Fanelli~D. \& Aurell~E., 2001
                submitted to Journ. Comp. Phys.,  cond-mat/0101336
  

\bibitem[1980]{Peebles1980}  Peebles~P.J. 1980, The Large-scale 
                Structure of the Universe, 
                (Princeton University Press, Princeton, NJ)

\bibitem[1992]{num}   Press,~H.W., Numerical Recipes in Fortran 1992,
                (Cambridge University Press, Cambridge) 


\bibitem[1990]{Rouet1990} Rouet~J.L., Feix~M.R. \& Navet M. 1990, 
                Vistas in Astronomy, 33, 357


\bibitem[1991]{Rouet1991} Rouet~J.L. et al 1991,
                in Lecture Notes in Physics:
                Applying Fractals in Astronomy, 161


\bibitem[1989]{Zeldovich} Shandarin~S.F. \& Zeldovich~Ya.B. 1989, 
                Rev. Mod. Phys., 61, 185

\bibitem[2000]{starobinsky} Starobinsky~A. 2000, Private Communication to 
U. Frisch 

\bibitem[1993]{Vergassola} Vergassola~M., Dubrulle~B., 
        Frisch~U. \& Noullez~A. 1993, A\&A, 289, 325. 

\bibitem[1972]{Weinberg1972} Weinberg~S. 1972, Gravitation and 
        Cosmology (Wiley)


\end{thebibliography}
\end{document}